\documentclass[slac_one]{revtex4}
\usepackage{graphicx}
\usepackage{fancyhdr}
\begin{document}

\title{{\small{2005 International Linear Collider Workshop - Stanford, U.S.A.}}\\ 

\vspace{12pt}
Beam Calorimeter Technologies} 

\author{R.~Dollan, Ch.~Grah, E.~Kouznetsova, W.~Lange, W.~Lohmann, A.~Stahl}
\affiliation{DESY, Zeuthen, Germany}

\author{K.~Afanaciev, V.~Drugakov, I.~Emeliantchik}
\affiliation{NCPHEP, Minsk, Belarus}

\begin{abstract}
Two different technologies are considered for the Beam Calorimeter of the ILC detector.
Simulation studies of the performance have been done for a diamond-tungsten sandwich 
calorimeter and for a homogeneous heavy element crystal calorimeter with optical fiber readout.
Studies of the stability and linearity of a polycrystalline diamond response were done
for the diamond-tungsten option. For the heavy crystal option the light yield reduction 
due to the wavelength shifting fiber readout was studied.
\end{abstract}

\maketitle

\thispagestyle{fancy}

\section{INTRODUCTION} 
The Beam Calorimeter (BeamCal) will be positioned just adjacent to the beampipe in front 
of the final focus quadrupoles covering very low angles (4-28)~mrad.
One of the purposes of the BeamCal is to serve fast beam diagnostics detecting $e^+e^-$ pairs 
originated from the beamstrahlung photon conversion. The calorimeter provides 
a good hermiticity of the whole detector and allows to measure 
high energy electrons down to the very low angles. The calorimeter also shields the inner part 
of the detector from backscattered beamstrahlung remnants and synchrotron radiation.

The beamstrahlung remnants create huge energy deposition in the BeamCal.  
The total energy deposited in the calorimeter is about 20~TeV 
per bunch crossing for nominal TESLA beam parameters and detector design. 
The deposited energy varies significantly with radius and azimuthal angle, 
providing areas of very high occupancy. The integrated radiation dose of up to 10~MGy/year 
is expected for these areas \cite{VFR1} requiring radiation hardness of the BeamCal 
sensitive material.

Measurements of high energetic electrons or photons on top of the
beamstrahlung remnants require a high transversal granularity and
a small Moliere radius of the calorimeter.

\begin{figure}
  \begin{minipage}{0.45\textwidth}
    \includegraphics[angle=270,width=8cm]{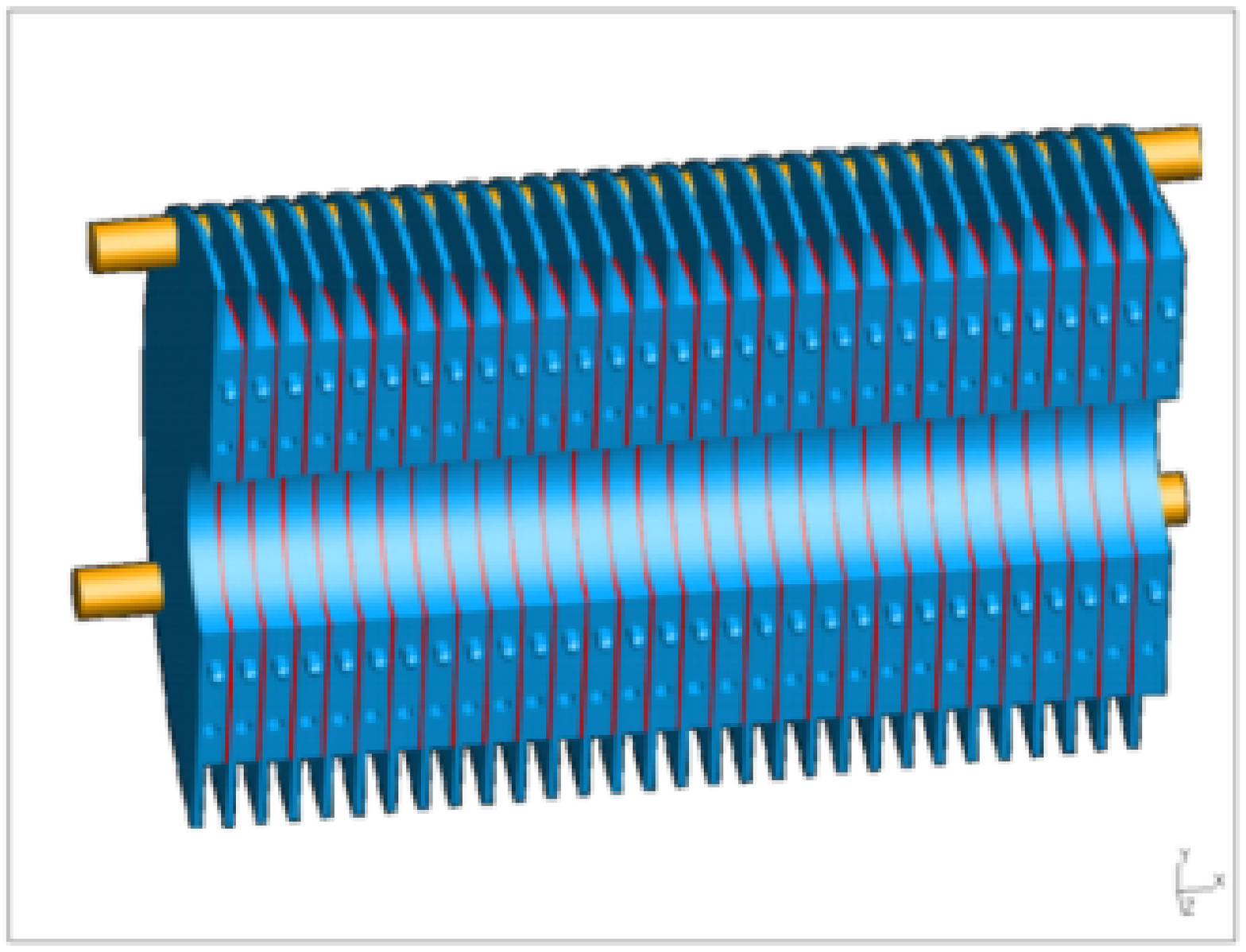}
    \caption[]{\label{fig:diamtung_sketch} The structure of a half barrel of the sandwich calorimeter. 
      Between the tungsten disks diamond sensors are interspersed.}
  \end{minipage}
  ~~
  \begin{minipage}{0.45\textwidth}
    \vspace{9mm}
    \includegraphics[width=8.0cm]{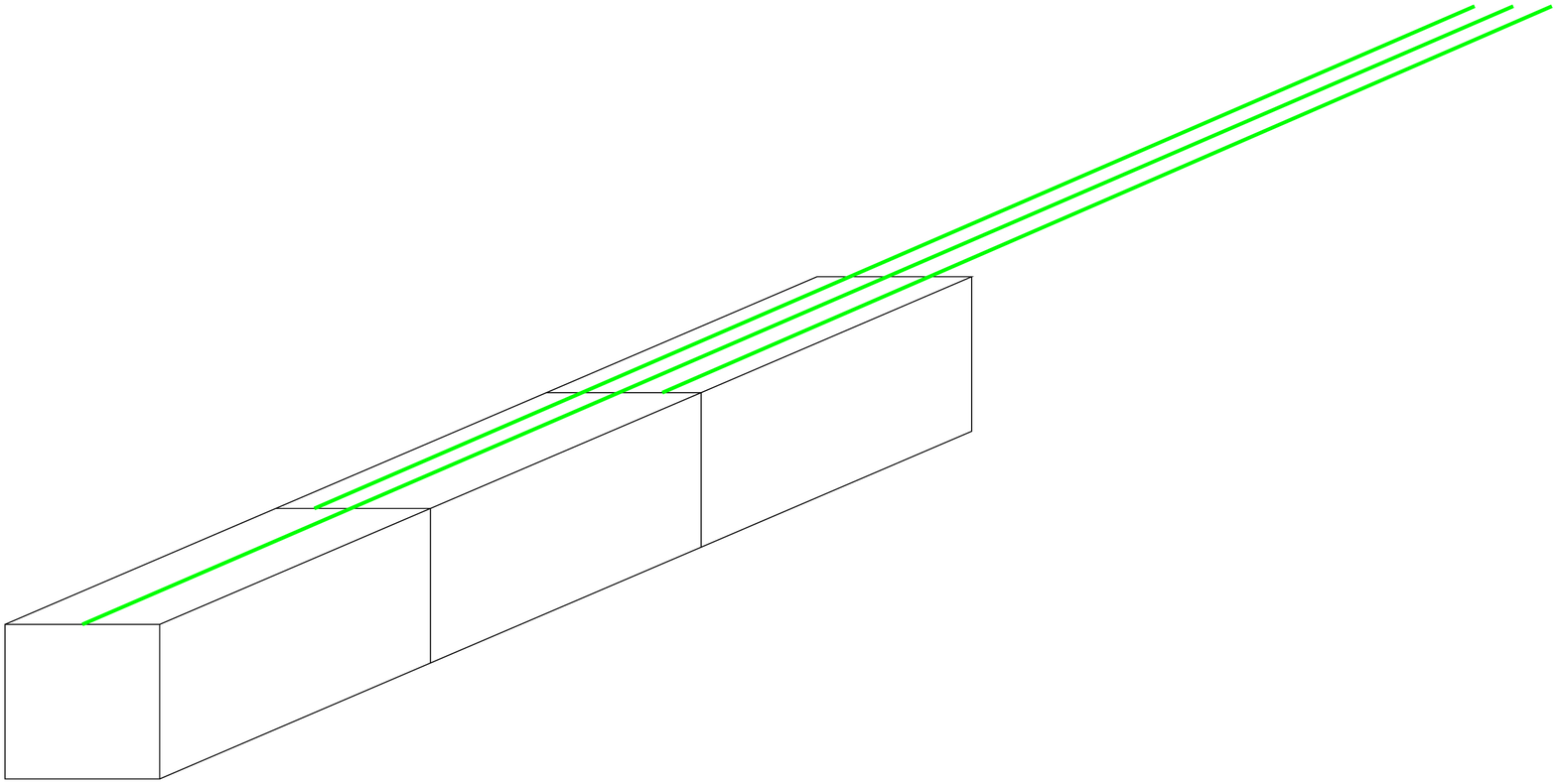}
    \vspace{1mm}
    \caption[]{\label{fig:scint_sketch} Scintillator pieces forming a longitudinal segment of 
      the crystal calorimeter. Each piece is connected to an optical fiber routed 
      to the back of the calorimeter.}
  \end{minipage}
\end{figure}

One of the considered technologies for the  BeamCal is a diamond-tungsten sandwich calorimeter. 
Polycrystalline Chemical Vapour Deposition (pCVD) diamonds have been shown to be sufficiently
radiation hard \cite{DIAM_behnke_2}. A sketch of a possible calorimeter structure is shown
in Figure~\ref{fig:diamtung_sketch}. 
 
Another technology considered for the BeamCal design is a heavy element homogeneous calorimeter
where scintillator segments are read out with optical fibers.
A longitudinal segment of the crystal calorimeter is shown in Figure~\ref{fig:scint_sketch}.
Every piece of the segment is optically isolated from the neighbors and read out with a
wavelength shifting fiber. The fiber is routed to the back of the calorimeter through optically 
isolated grooves in the adjacent rear parts. As a possible material for the calorimeter lead 
tungstenate (PbWO$_4$) is considered in the simulations.

\section{SIMULATION OF THE CALORIMETER PERFORMANCE} 
The beamstrahlung is simulated using the Monte Carlo program Guinea Pig~\cite{gpig}. 
Single high energy electrons are simulated with energies 
between 50 and 250 GeV using the GEANT3~\cite{geant3} based detector simulation
package BRAHMS~\cite{brahms}.

The reconstruction procedure was tuned to provide less than 1\% of fake events.
The fake rate was estimated by applying the reconstruction algorithm to pure 
background events. 

\subsection{Diamond-Tungsten Calorimeter}   %
The simulated calorimeter consists of 30 tungsten disks alternating with diamond 
sensor layers. The thickness of the tungsten disks is chosen to be 3.5~mm (one radiation length),
the diamond layers are 0.5~mm thick. Every diamond layer is segmented into pads.
The number of pads in a ring increases with the radius keeping pad dimensions of about half 
a Moliere radius (5~mm). 

\begin{figure}[htb]
\begin{minipage}{0.45\textwidth}
\includegraphics[width=7.5cm]{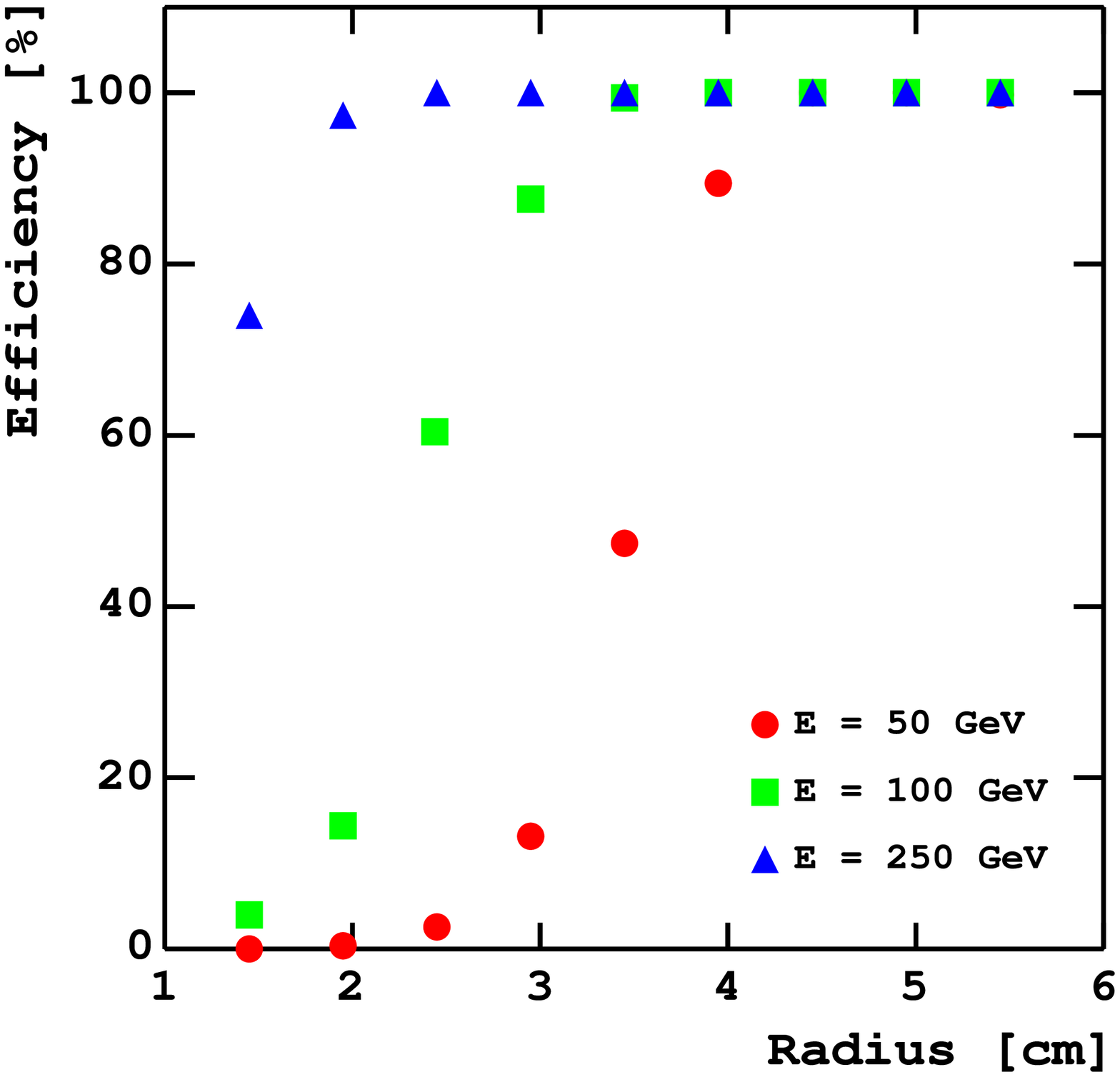}
\caption[]{\label{fig:diamtung_eff1} The efficiency to detect electrons of 50, 100 and 250~GeV in the
  high background region ($\phi\approx 90^o$) of the BeamCal as a function of the radius of the calorimeter.}
\end{minipage}
\begin{minipage}{0.02\textwidth}
~~
\end{minipage}
\begin{minipage}{0.45\textwidth}
\includegraphics[width=8.5cm]{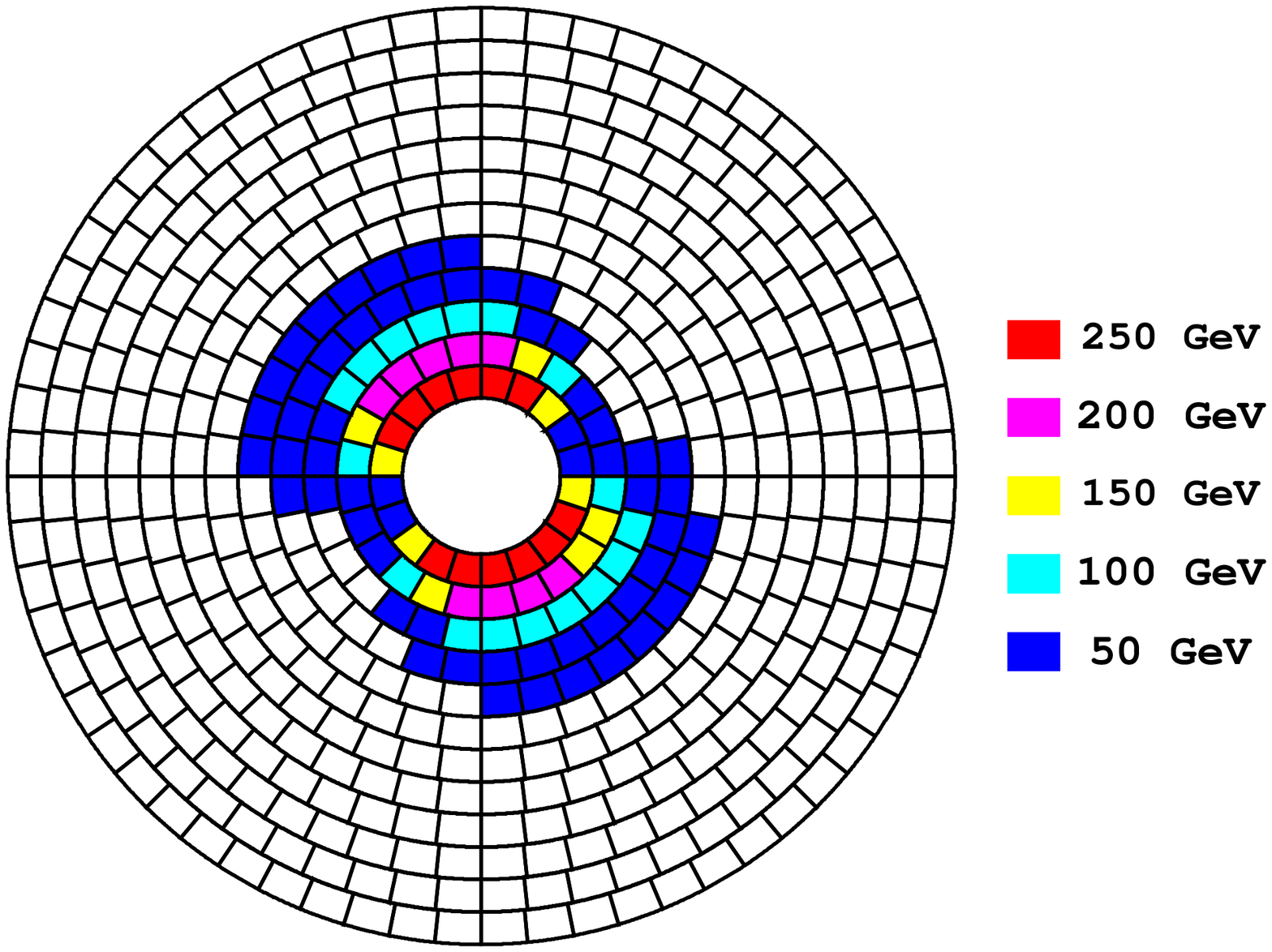}
\caption[]{\label{fig:diamtung_eff2}  A map of low-efficiency regions. The cells where the detection
  efficiency for a high energetic electron is less than 90\% are marked with colors 
  corresponding to the electron energy.}
\end{minipage}
\end{figure}

Figure~\ref{fig:diamtung_eff1} shows the detection efficiency as a function of the radius 
for electrons of different energies. The efficiencies are obtained 
at azimuthal angle around $\phi= 90^o$ where the background level is  high. 
Figure~\ref{fig:diamtung_eff2} shows a map of cells where the detection efficiency is less than 90\%.
Electron energies between 50 and 250~GeV are considered.

\subsection{Heavy Scintillator Calorimeter} %
\begin{figure}[htb]
\begin{minipage}{0.45\textwidth}
\includegraphics[width=7cm]{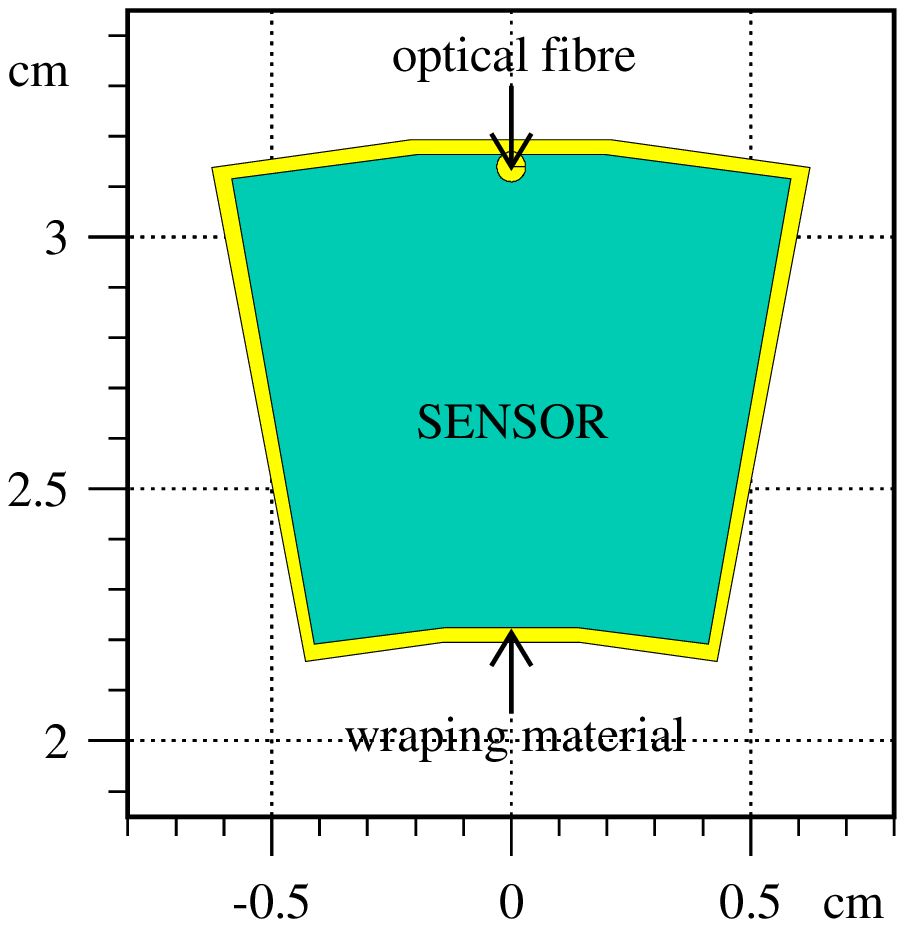}
\caption[]{\label{fig:leadtung_segm} A cross-section of a frontal calorimeter segment 
  with wrapping and wavelength shifting fiber.}
\end{minipage}
~~
\begin{minipage}{0.45\textwidth}
\includegraphics[width=8cm]{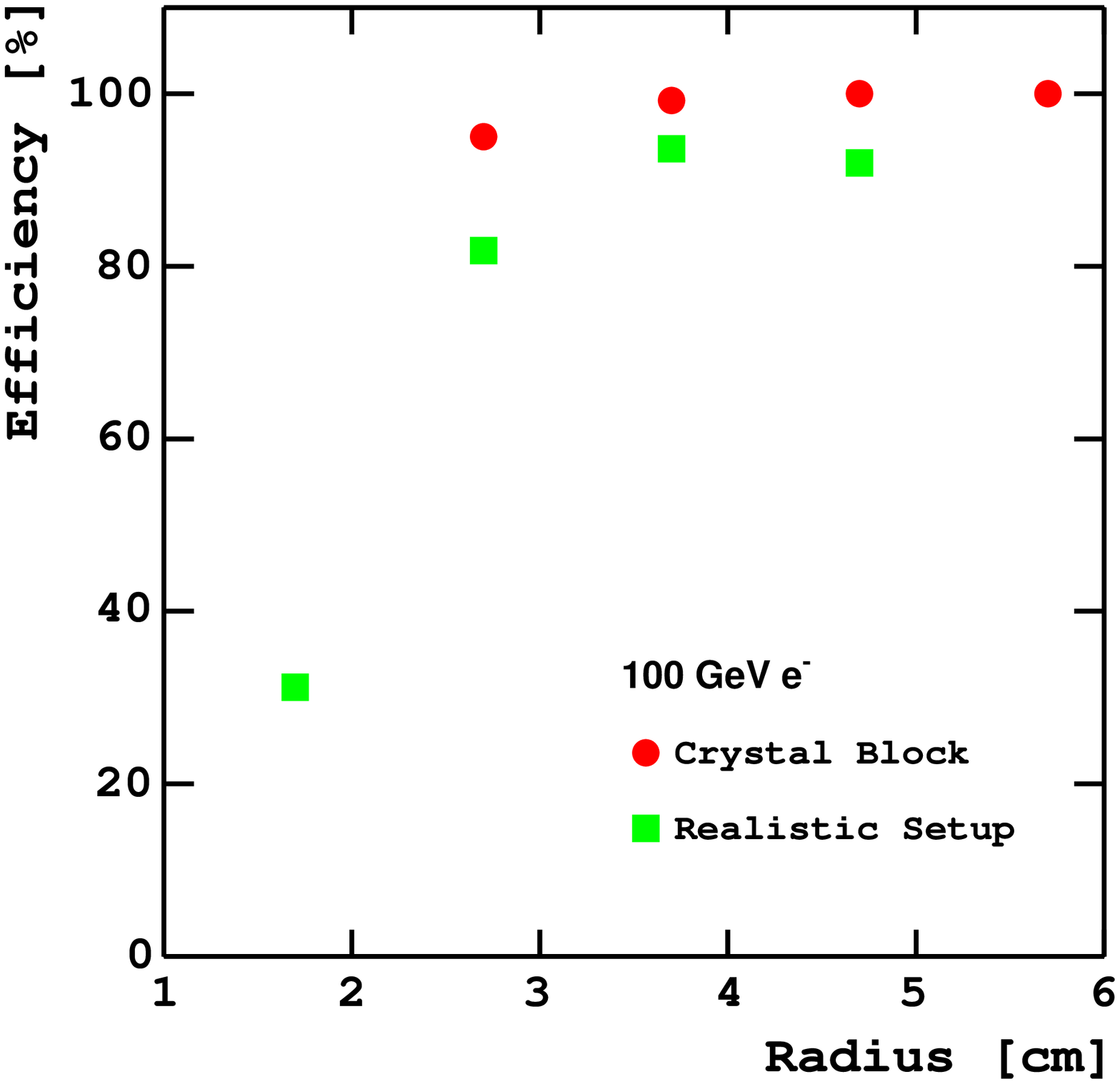}
\caption[]{\label{fig:leadtung_eff} The detection efficiency for 100~GeV electrons 
  in the low background region as a function of the radius. The results for the ideal 
  (circles) and realistic calorimeter with fibers and wrapping material (squares) are shown.}
\end{minipage}
\end{figure}

The study of a lead tungstenate calorimeter performance is done the same way.
The transverse segmentation is chosen to be of about half a Moliere radius
($\sim1$~cm). Every segment is divided longitudinally into tree pieces as it is shown 
in Figure~\ref{fig:scint_sketch}. The length of the pieces is 3, 9 and 8 radiation 
lengths starting with the front side of the calorimeter. To investigate the influence of 
the fibers and wrapping material on the calorimeter performance, an ideal homogeneous
calorimeters has been studied as well as a realistic one. Figure~\ref{fig:leadtung_segm} shows
a cross-section of a frontal calorimeter segment with wrapping and an attached fiber
as it is implemented for the realistic simulation.

Figure~\ref{fig:diamtung_eff1} shows the efficiency to detect a 100~GeV electron in the low background region
at azimuthal angle around $\phi= 0^o$ as a function of the radius. The deterioration of the performance due to
the wrapping and fiber material is clearly seen.

\section{SENSOR STUDIES FOR DIAMOND-TUNGSTEN OPTION} 
Polycrystalline CVD diamond samples produced at Fraunhofer Institut f\"ur Angewandte Festkörperphysik 
(Fraunhofer~IAF, Freiburg, Germany)
have been tested for the diamond-tungsten option of the BeamCal. 
The samples have 10x10~mm$^2$ area with a two-layer Ti/Au metalization and
thickness between 200 and 500~$\mu$m. 

Measurements of the charge collection efficiency under low irradiation doses 
have been done in order to check stability of the signal. 
A diamond response to irradiation with a $^{90}$Sr beta-source was monitored 
up to an absorbed dose of about 20~Gy. The voltage applied during 
the irradiation corresponded to an electric field of 1~V/$\mu$m.

\begin{figure}[htb]
\begin{minipage}{0.45\textwidth}
\vspace{3mm}
\includegraphics[width=7cm,height=8cm]{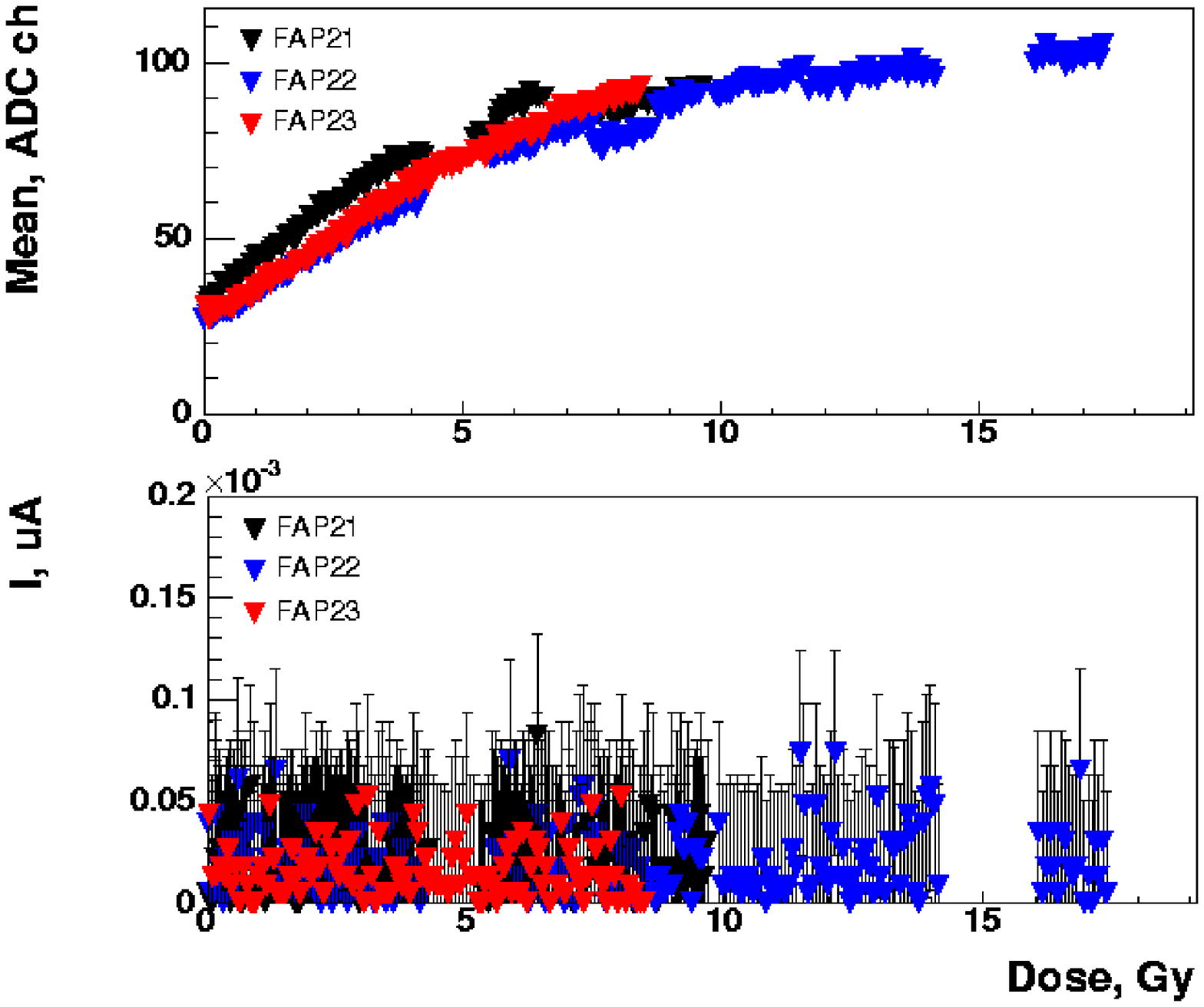}
\caption[]{\label{fig:diamond_dose1} The  mean values of the $^{90}$Sr spectrum
  (top) and current in the high voltage circuit (bottom) obtained for three samples 
  as function of the absorbed dose.}
\end{minipage}
~~
\begin{minipage}{0.45\textwidth}
\includegraphics[width=8cm, height=8.5cm]{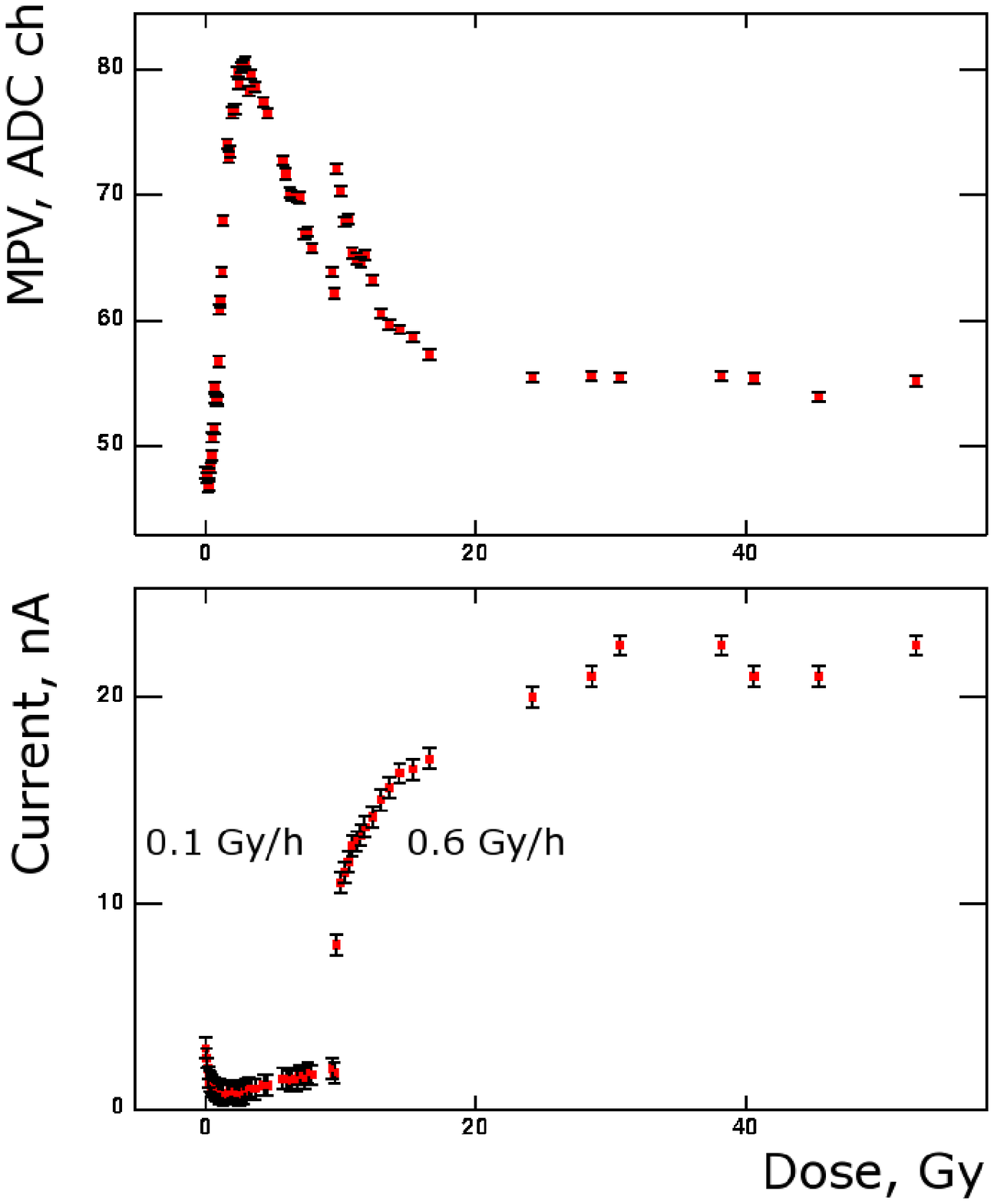}
\caption[]{\label{fig:diamond_dose2} The  mean values of the $^{90}$Sr spectrum 
  (top) and current in the high voltage circuit (bottom) as function of the absorbed dose. 
  The dose accumulation rate was changed at about 10~Gy.}
\end{minipage}
\end{figure}

The top plot of Figure~\ref{fig:diamond_dose1} shows the mean values of the $^{90}$Sr spectrum
obtained with three different diamond samples as a function of the absorbed dose. The charge 
collection efficiency stabilizes after the dose of about 15~Gy.
Stability of current in the high voltage circuit during the irradiation
is shown in the bottom plot of Figure~\ref{fig:diamond_dose1}.

Figure~\ref{fig:diamond_dose2} shows results obtained with another diamond sample.
The current increases with the absorbed dose and depends on the dose rate.
Stabilization of the current and charge collection efficiency has been obtained at 
about 20~Gy.

\begin{figure}[htb]
\includegraphics[width=10cm]{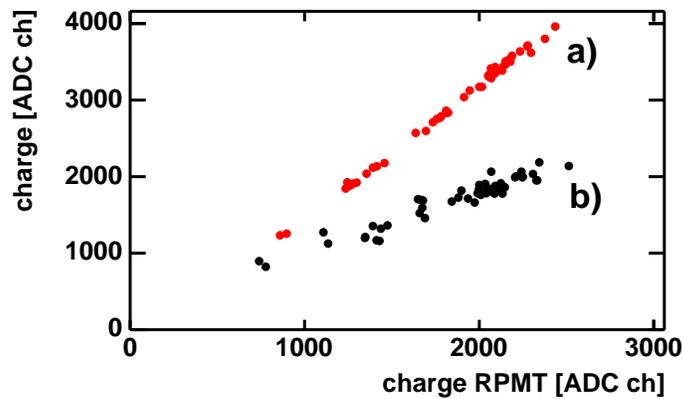}
\caption[]{\label{fig:diamond_linear} The signal from a diamond (a) and one of the
photomultipliers (b) as a function of the signal from the second photomultiplier.}
\end{figure}

The linearity of a diamond response has been tested with a hadron beam of about 4~GeV at 
the CERN~PS. Beam spills of about 10~ns duration with variable intensity (up to $10^7$ 
mips per spill) produced an integrated signal in a diamond sample. A scintillator with
two photomultipliers attached was used as a trigger and as a reference for the
linearity measurements. An example of the measurement results is shown in 
Figure~\ref{fig:diamond_linear}.

\section{STUDIES OF FIBER READOUT FOR CRYSTAL OPTION} 
\subsection{Light Yield Reduction and Crosstalk Measurements}
\begin{figure}[h]
  \centerline{\hbox{ 
      \includegraphics[width=6cm]{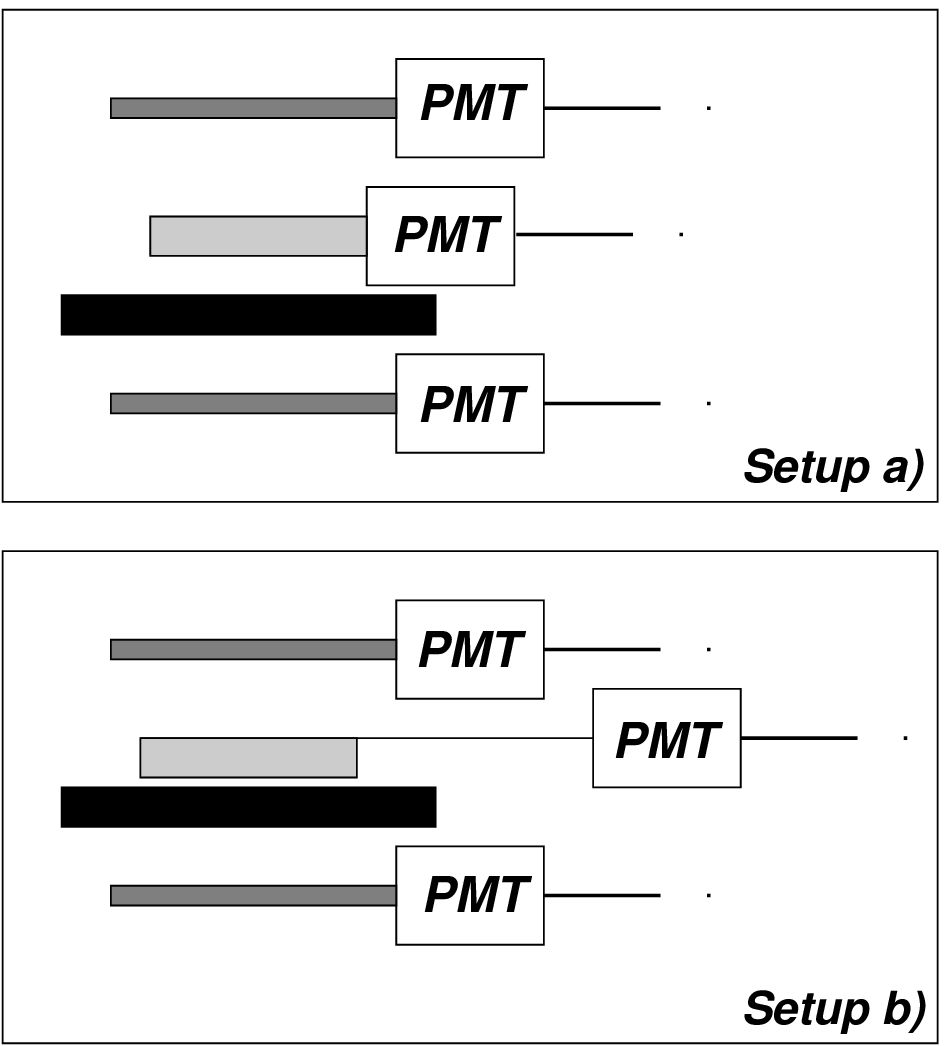}
      \includegraphics[width=8.2cm]{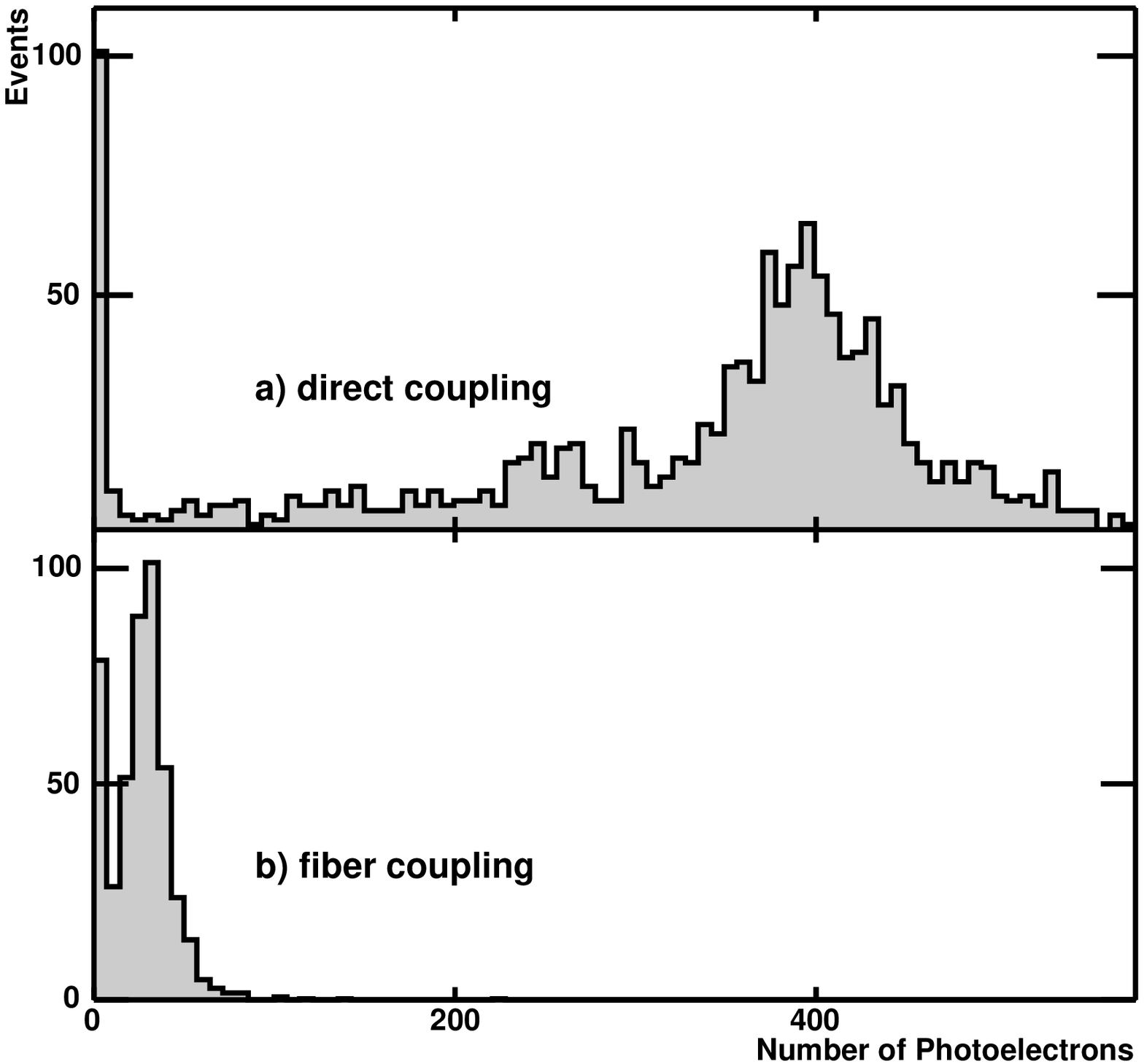}
    }
  }
\caption[]{\label{fig:scint_setup} Setup for light yield measurements (left). The scintillator
  under the test is positioned between triggering scintillators and read out directly  (Setup "a") 
  or via a  wavelength shifting fiber (Setup "b"). The signal spectra obtained from relativistic cosmic 
  muons for Setup~"a" and Setup~"b" (Right).}
\end{figure}

Measurements of light yield reduction due to the fiber readout 
have been done in order to test the feasibility of the concept.
The measurements are done using plastic scintillator (Bicron BC-408~\cite{bicron}) 
and lead glass as the test samples. 
These two materials have different light production mechanisms: in plastic 
scintillator light is produced due to scintillation, in lead glass 
via the \v Cerenkov effect.
Light produced in a crystal under test due to 
traversing cosmic muons is read out by a photomultiplying tube
coupled to the scintillator directly (Setup "a" in Figure~\ref{fig:scint_setup},~left)
or via a Bicron BCF-19A wavelength shifting fiber (Setup "b" in 
Figure~\ref{fig:scint_setup},~left). An example of the obtained spectra is shown in 
Figure~\ref{fig:scint_setup},~right.
The light yield obtained with the fiber readout method compared to a direct 
readout is reduced to (14$\pm$4)\% for plastic scintillator 
and (16$\pm$7)\% for lead glass.

To estimate the crosstalk for the considered readout, measurements 
with a plastic scintillator segmented into three pieces were done.
Every segment had one readout fiber attached and wrapping for optical insulation.
The fibers were routed to one side as it is shown in Figure~\ref{fig:scint_sketch} 
and were optically isolated from the neighboring segments.

All three fibers were read out with photomultipliers. Setup "b" shown in 
Figure~\ref{fig:scint_setup} was rearranged in a way to produce a trigger signal
only for muons traversing the first segment, where all three fibers were routed through.
The signals from the other two fibers were monitored. 
The crosstalk level was measured to be less than 1\%.

\subsection{Simulation Studies of Light Yield Reduction}
\begin{figure}[htb]
\includegraphics[width=7.5cm]{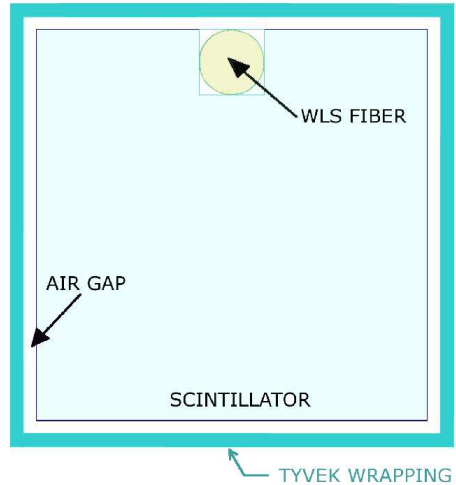}
\caption[]{\label{fig:ctystal_sim_geom} A cross-section of the scintillator
considered in the simulation. The scintillator has a rectangular slot with
a glued wavelength shifting fiber.}
\end{figure}

The transport of optical photons generated in a crystal was
simulated using GEANT4~\cite{geant4}. Scintillation and \v Cerenkov radiation as well
as a set of optical processes including the wavelength shifting mechanism were taken into account.
The wavelength shifting process is available since version~6.0 of GEANT4.
The simulation is done for the plastic scintillator and lead glass crystals
of the same geometry as in the measurements of light yield reduction. 
A cross-section of the crystal is shown in Figure~\ref{fig:ctystal_sim_geom}. 
The wavelength shifting fiber and its connection to the crystal were thoroughly implemented 
including the fiber material, fiber cladding and optical glue.  The wavelength shifting process 
was based on the absorbtion and emission spectra of BCF-19A fibers provided by Bicron.
For the simulation no optimization of the material surfaces and  the material boundary
conditions were done.

A light yield reduction to (9.3-9.8)\% for the plastic scintillator 
and (8.3-12.0)\% for the lead glass has been obtained from the simulation, 
that is in good agreement with the results of the measurement.

\section{CONCLUSION} 
The simulation studies of the BeamCal performance showed the considered technologies 
to be feasible. The efficiency to detect electrons with energy of about 
100~GeV and higher is almost 100\% for most of the calorimeter area. 

The ongoing studies of diamond samples for the diamond-tungsten options are done
in contact with the diamond manufacturer (Fraunhofer~IAF). The goal of the work is 
to prepare a calorimeter prototype with reliable diamond sensors for further 
studies in a test-beam.

The fiber readout for the heavy crystal calorimeter is shown to be feasible. The
GEANT4 simulation of the light transportation is shown to be in good agreement
with the measurement results. This allows it to be included into the realistic simulation of the
heavy crystal BeamCal.

\end{document}